\documentclass[12pt]{article}
\usepackage{graphicx}

\begin{document}

\begin{center}
{\Large \bf
 Einstein's Hydrogen Atom}

\vspace{2ex}

Y. S. Kim \footnote{electronic address: yskim@umd.edu}\\
Department of Physics, University of Maryland,\\
College Park, Maryland 20742, U.S.A.\\

\end{center}

\vspace{2ex}

\begin{abstract}
In 1905, Einstein formulated his special relativity
for point particles.  For those particles, his Lorentz covariance
and energy-momentum relation are by now firmly established.  How
about the hydrogen atom?  It is possible to perform Lorentz
boosts on the proton assuming that it is a point particle.
Then what happens to the electron orbit?  The orbit could go
through an elliptic deformation, but it is not possible to
understand this problem without quantum mechanics, where
the orbit is a standing wave leading to a localized probability
distribution.  Is this concept consistent with Einstein's Lorentz
covariance?  Dirac, Wigner, and Feynman contributed important
building blocks for understanding this problem.  The remaining
problem is to assemble those blocks to construct a Lorentz-covariant
picture of quantum bound states based on standing waves. It is
shown possible to assemble those building blocks using harmonic
oscillators.
\par

\end{abstract}

\newpage
\section{Introduction}\label{intro}
Niels Bohr had a great respect for Einstein, and he adds ``time''
whenever he mentions ``space'' in his philosophical writings.
However, for his hydrogen atom, the proton was sitting at the
center of the absolute frame.  Einstein presumably thought about
how the hydrogen atom would look to a moving observer, but he never
raised the issue.   The reason is that the hydrogen atom moving
with a relativistic speed was not conceivable for them.
\par
\begin{figure}[thb]
\centerline{\includegraphics[scale=0.3]{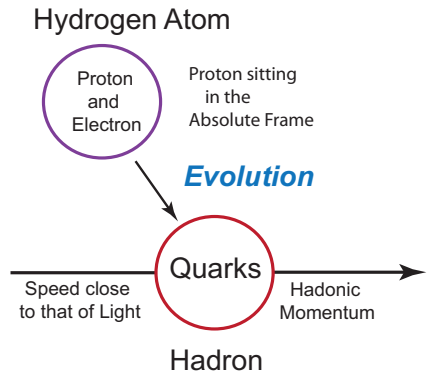}}
\caption{Evolution of the hydrogen atom.  It is still not possible to
accelerate the hydrogen atom to a relativistic speed.  It is however
possible these days to produce protons moving with a speed close to
the light speed.  Also these days, the proton is thought to be a bound
state of quarks. It is thus sufficient to study high-energy protons to
study the hydrogen atom in Einstein's world.}\label{evol33}
\end{figure}

Things are different these days.  Protons can move with a speed
close to the light speed.  In addition, like the hydrogen atom,
the proton is a bound state of the more fundamental particles
called the ``quarks."  The proton thus has the same quantum
mechanical ingredients as the hydrogen atom has.  We can therefore
study the hydrogen atom in Einstein's world by studying the proton
in high-energy physics. This historical trend is illustrated in
Fig.~\ref{evol33}.

   \par
Without the quark model, Paul A. M. Dirac devoted much of research
life to the problem of constructing Lorentz-covariant wave functions.
He published four papers on this problem from 1927 to
1963~[1--4].  We shall construct the
bound-state model by combining these four papers.
\par
In order to do this, we have to understand the symmetry problems for
bound-state problems.  In 1939, Eugene Wigner worked out the internal
space-time symmetries of relativistic particles~[5].  In
so doing he worked out the symmetries of bound states in the
Lorentz-covariant world~[6].
\par
Richard Feynman invented Feynman diagrams, but he said in 1970 that
we should use harmonic oscillators, instead of Feynman diagrams, for
understanding bound state problems in the Lorentz-covariant
world~[7].  He then published a paper saying the same
with his students in 1971~[8].

\par
In Sec.~\ref{dirac4}, we list Dirac's four papers, and point out what
he did and what he could have don in these papers.  We do the same
for Feynman's three papers in Sec.~\ref{feyn3}.
In Sec~\ref{covham}, it was noted first that space-time symmetry of
quantum bound states is simpler than the full-fledged Lorentz group.
Unlike Klein-Gordon waves, the symmetry of standing wave is that of
the three-dimensional rotation group~[5].  This point is
missing in Dirac's papers and Feynman's 1971 paper~[8].  It
is noted also that that the covariant harmonic oscillators satisfy
all the required symmetries.
\par
We then discuss the essential features of the oscillator formalism
which describes the effect of the proton wave function under
Lorentz boot.  It is shown that the wave function becomes ``squeezed''
when it is boosted.
\par
It is then shown in Sec.~\ref{fparton} that this squeeze effect
manifests itself in Feynman's parton picture for the proton moving
with a speed close to that of light. We establish that the quark
model and the parton model are two different manifestations of one
Lorentz-covariant model of quantum bound states.  This is what
Einstein's hydrogen atom is about.

\vspace{-2.0mm}
\section{Dirac's Four Papers}\label{dirac4}

\vspace{-2.0mm}
Paul A, M. Dirac devoted much of his research efforts to making
quantum mechanics consistent with special relativity.

\begin{itemize}

\item  In his 1927 paper on time-energy uncertainty
relation~[1], Dirac noted that there are no quantum
excitations along the time variable, unlike Heisenberg's
position-momentum relation.  He said this space-time asymmetry
makes the problem difficult.

\item  In 1945~[2], Dirac attempted to construct harmonic
oscillator wave functions which can be Lorentz-boosted.  He wrote
down the Gaussian form
\begin{equation}\label{gauss11}
\exp{\left[  -\frac{1}{2}\left(x^2 + y^2 + z^2 + t^2\right)\right] } ,
\end{equation}
but did not explain the physics of the Gaussian distribution in the
time variable.

\item  In 1949~[3], he started with the  Lorentz transformation
\begin{equation}\label{boost11}
\pmatrix{z' \cr t'} = \pmatrix{\cosh \eta & \sinh \eta \cr
\sinh \eta & \cosh \eta } \pmatrix{z \cr t}.
\end{equation}
He then introduced the light-cone variables
\begin{equation}\label{boost33}
  u = \frac{z + t}{\sqrt{2}}, \qquad v = \frac{z - t}{\sqrt{2}} .
\end{equation}
In terms of these variables, the Lorentz boost takes the form
He then diagonalize this equation to
\begin{equation}\label{boost22}
u'  = e^\eta u, \qquad v' = e^{-\eta} v .
\end{equation}
These light-cone variables serve very useful purposes.  Here one
coordinate expands and the other contracts.  Thus, the Lorentz
boost is a squeeze transformation~[9].
\par
In the same paper, Dirac stated that the problem of constructing
relativistic dynamics is the same as that of constructing a
suitable representation of the Poincar\'e group.  In his earlier
paper~[2], Dirac started this work using harmonic
oscillators, but he did not elaborate on this in his 1949 paper.

\item In 1963~[4], Dirac used two harmonic oscillators
 to construct the $O(3,2)$ deSitter group, which is a Lorentz group
 applicable to thee space-like and two time-like coordinates.
 This representation later became the mathematical basis for
 two-mode squeezed states in quantum optics~[10,11],
 and became a bridge between special relativity and optical
 sciences.
\end{itemize}

In the present paper, we address these soft spots in these papers
according to Dirac's own suggestion: to construct the representation
of the Poincar\'e group using harmonic oscillators~[2,3].
Dirac missed this point again in his 1963 paper~[4] while
he was constructing the representation of the $O(3,2)$ group which
contains the Lorentz group $O(3,1)$ as a subgroup.

\par
We can remove these soft spots by constructing Wigner's little
groups~[5] of the Poincar\'e group using harmonic
oscillators~[6,12].

\vspace{-2.0mm}
\section{Feynman's Three Papers}\label{feyn3}

\vspace{-2.0mm}
Richard Feynman made important contributions in many different
branches of physics.  In the following three papers, he left
some important questions as home work problems for younger
generations.

\begin{itemize}

\item  In 1969~[13,14], Feynman introduced the
   concept of partons.  If the proton moves with a velocity close
   to that of light, it appears like a collection of partons whose
   properties are quite different from the quarks which are constituent
   particles inside the proton at rest.  The question then is
   whether the quarks and partons are two different manifestation
   of one Lorentz-covariant entity.

\item  In 1970, Feynman gave a talk at the spring meeting of the
   American physical Society held in Washington.  He started with
   hadrons which are bound states of quark~[15].   He noted
   that the hadronic spectra could best be understood in terms
   of the three-dimensional harmonic oscillators. As for the Lorentz
   covariant aspect of his oscillator formalism, he pointed out
   that there is the time separation between the quarks.  However,
   since he did not know what to do with it, he chose to ignore
   the variable.  He then published the content of this talk with
   his students in 1971~[8].   He did not justify what he
   did on this time separation variable.

\item  In his book on statistical mechanics published in
  1972~[16],  Feynman discussed density matrices and
  measurement problems.  He stated
  {\it When we solve a quantum-mechanical problem, what we really do
  is divide the universe into two parts - the system in which we are
  interested and the rest of the universe.  We then usually act as if
  the system in which we are interested comprised the entire universe.
  To motivate the use of density matrices, let us see what happens
  when we include the part of the universe outside the system.}
\par
  Feynman then used one harmonic oscillator to illustrate his rest
  of the universe.  The question is how one oscillator can explain
  both the real world and the rest of the universe.  He could have
  used two coupled oscillators to illustrate his his rest of
  the universe, but he left this problem as a homework problem for
  us~[17].
\end{itemize}
\par
In these three papers, Feynman raised very fundamental issues in
physics, but did not provide complete solutions.  The issue on
his rest of the universe has been discussed in the literature
in terms of the coupled oscillators~[17], and also
in terms of the time-separation variable in the Lorentz-covariant
world~[18].

In the present paper, we are interested in addressing the soft spots
in Feynman's 1969 papers on the parton picture and those in his 1971
paper on harmonic oscillators.  As in the case of Dirac, it is
possible to transform Feynman's oscillator formalism into the
representation of Wigner's little group using harmonic
oscillators~[6,12].

\vspace{-2.0mm}
\section{Covariant Harmonic Oscillators}\label{covham}

\vspace{-2.0mm}
In Sec.~\ref{dirac4} and Sec.~\ref{feyn3}, we stated that it is
possible to remove the soft spots in Dirac's four papers and
Feynman's three papers by constructing Wigner's little groups.
These little groups are the subgroups of the Poincar\'e group
whose transformations leave the four-momentum of a given
particle invariant~[5,6].  For a massive particle,
we can consider the Lorentz frame where this particle is at
rest.  In this frame, the space-time symmetry is the three-dimensional
rotation group.
\par
\begin{figure}[thb]
\centerline{\includegraphics[scale=0.4]{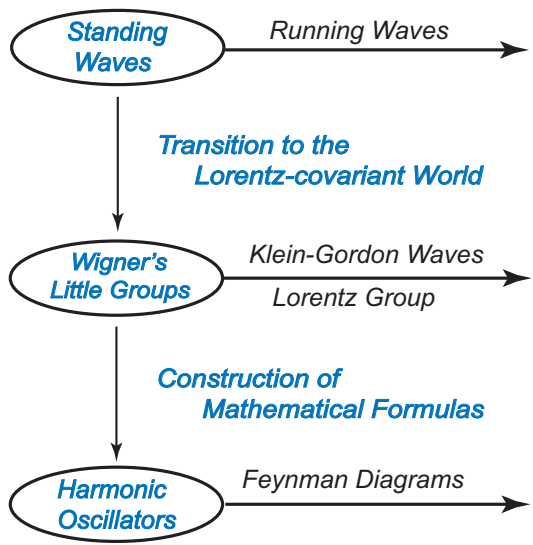}}
\vspace{5mm}
\caption{Running waves and standing waves in quantum theory.  If a
particle is allowed to travel from infinity to infinity, it corresponds
to a running wave according to the wave picture of quantum mechanics.
If, on the other hand, it is trapped in a localized region, we have
to use standing waves to interpret its location in terms of
probability distribution.}\label{standrun22}
\end{figure}

In dealing with plane waves, we start with the Klein-Gordon equation.
The solutions of this equation are Lorentz-invariant.  The running
waves in the Lorentz-covariant world share the same symmetry property
as that of the Klein-Gordon waves.  It contains the full symmetry of
the Poincar\'e group with ten interdependent paperers.
These aspects of the space-time symmetry is illustrated in
Fig.~\ref{standrun22}.  This figure describes the space-time
symmetry of Einstein's hydrogen atom given in Fig.~\ref{evol33}.

Since the internal space-time symmetry is like the three-dimensional
rotation group, the standing waves trapped within a quantum bound
state should also satisfy this symmetry.  It is important to note
that we are dealing here with space-time separations.  For instance,
the Bohr radius is the separation between the proton and electron.
One of the soft spots in Dirac's four papers is that Dirac did not
clarify this separation issue.  The soft spots in both Dirac's papers
and Feynman's 1971 paper~[8] is that the time-like direction
is not required in Wigner's three-dimensional space.
\par
Thus, Dirac's concern about the space-time asymmetry is not necessary.
Feynman {\it et al.} said they wanted to drop the time-like
variable because they do not know what to do with it.  They did not
know they were right.  They did not have to do anything about what
does not exist.
\par
Then, our next problem is to build a model of bound states satisfying
Wigner's $O(3)$-like symmetry, which is consistent with Einstein's
Lorentz covariance.  As was noted by Feynman~[7],
the easiest way is to start with harmonic oscillators.  The
oscillator system does not require additional boundary conditions.
Indeed, before the paper of Feynman {\it et al.}, a number of
authors published their papers on this
subject~[19--23].
\par

According to Gell-Mann~[15], the proton is a bound state of
two quarks, but we consider here for simplicity a bound state of two
quarks.  As is the case of Feynman {\it et al.}, we start with
the two quarks whose  space-time positions are $x_{a}$ and $x_{b}$.
Then the standard procedure is to use the variables
\begin{equation}
X = (x_{a} + x_{b})/2 , \qquad x = (x_{a} - x_{b})/2\sqrt{2} .
\end{equation}
The four-vector $X$ specifies where the proton is located in space and
time, while the variable $x$ measures the space-time separation
between the quarks.  This $x$ variable has four components, but it has
only three degrees of freedom according to Wigner's symmetry.  This
will appear as the lack of excitations along the time-like direction
as noted by Dirac~[1,3].

\par
Does this time-separation variable exist when the proton is at rest?
Yes, according to Einstein.  In the present form of quantum mechanics,
we pretend not to know anything about this variable.  Indeed, this
variable belongs to Feynman's rest of the universe~[18].
\par
Also in the present form of quantum mechanics, there is an uncertainty
relation between the time and energy variables.  However, there are
no known time-like excitations.  Unlike the position or momentum variable,
the time-separation variable is c-number, and its uncertainty with the
energy separation a c-number uncertainty relation~[1].  With
this point in mind, let us go to the oscillator formalism proposed
by Feynman~[7,8].

\par
Feynman {\it et al.} start with the Lorentz-invariant differential
equation~[8]
\begin{equation}\label{osceq}
{1\over 2} \left\{x^{2}_{\mu} - {\partial^{2} \over \partial x_{\mu }^{2}}
\right\} \psi(x) = \lambda \psi(x) .
\end{equation}
This partial differential equation has many different solutions
depending on the choice of separable variables and boundary conditions.
Feynman {\it et al.} insist on Lorentz-invariant solutions which are
not normalizable.  On the other hand, if we insist on normalization,
the ground-state wave function takes the form of Eq.(\ref{gauss11}),
which now can be written as
\begin{equation}\label{gauss22}
\psi(z,t) = \exp{\left[ - \frac{1}{2}\left(z^2 +
                  t^2\right)\right] } ,
\end{equation}
where we dropped the transverse components of $x$ and $y$.  As in
the case of Eq.(\ref{boost11}), we make Lorentz boosts along the
$z.$  We dropped also the normalization constant for simplicity.
In terms
of the light-cone variables, this wave function becomes
\begin{equation}\label{gauss33}
\psi{u,v} = \exp{\left[ - \frac{1}{2}\left(u^2 +
                  v^2\right)\right] } .
\end{equation}

\par
\begin{figure}[thb]
\centerline{\includegraphics[scale=0.35]{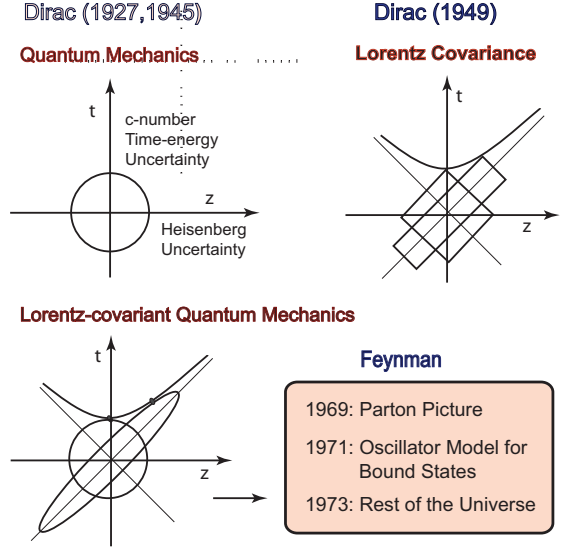}}
\vspace{5mm}
\caption{Space-time picture of quantum mechanics.  There
are quantum excitations along the space-like longitudinal direction, but
there are no excitations along the time-like direction.  The time-energy
relation is a c-number uncertainty relation.}\label{dirackn33}
\end{figure}

If the system is boosted, the $u$ and $v$ variables are replaced by
$u~e^{-\eta}$ and $v~e^{\eta}$ respectively.  The wave function then
becomes
\begin{eqnarray}\label{gauss44}
&{}& \exp{\left[ - \frac{1}{2}\left(e^{-2\eta} u^2 +
                  e^{2\eta} v^2\right)\right] } \nonumber \\[1.0ex]
&{}&\hspace{3mm} = \exp{\left[ -\frac{1}{4}\left(e^{-2\eta} (z + t)^2 +
                    e^{2\eta}(z - t)^2\right)\right] } .
\end{eqnarray}
The wave function satisfied the Lorentz-invariant differential equation
of Eq.(\ref{osceq}).  This wave function is expanded along the u direction,
while it becomes contracted along the $v$ direction.  This aspect of the
Lorentz-squeeze is illustrated in Fig.~\ref{dirackn33}.

\par

Let us go back to the Gaussian form of Eq.(\ref{gauss22}).  If we allow
excitations along the $z$ direction while keeping the $t$ component in
its ground state, the wave function takes the form
\begin{equation}\label{gauss55}
\psi_{n}(z,t) = \exp{\left[ - \frac{1}{2}\left(z^2 +
                  t^2\right)\right] } H_{n}(z) ,
\end{equation}
where $H_{n}$ is a Hermite polynomial.  This wave function satisfies Dirac's
 c-number time-energy uncertainty relation.  It can also be Lorentz-boosted
in the same manner as the ground-sate wave function of Eq.(\ref{gauss22})
becomes its squeezed form of Eq.(\ref{gauss44}).  This aspect of the
Lorentz-covariant c-number time-energy uncertainty relation is discussed
in the literature~[24,25].
\par
Since the oscillator system is separable in the Cartesian coordinate
system, the Gaussian form of Eq.(\ref{gauss22}) can be restored to its
dimensional form of Eq.(\ref{gauss11}).  This form can allow excitations
along the transverse directions of $x$ and $y$.  We we add the Hermite
polynomials in along these components, this wave function can possess
the symmetry under rotations in the three-dimensional space.  This is
the content of Wigner's $O(3)$-like little group applicable to this
system.  These transverse excitation remain invariant when the
system is boosted.  This aspect has also been discussed in the
literature~[12].

\vspace{-2.0mm}

\section{Feynman's Parton Picture}\label{fparton}

\vspace{-2.0mm}
It is a widely accepted view that the hadrons are quantum
bound states of quarks with localized probability distributions.
As in all bound-state cases, this localization condition is
responsible for the existence of discrete mass spectra.  The most
convincing evidence for this bound-state picture is the hadronic
mass spectra which are observed in high-energy
laboratories~[6,8].  The proton is one of those hadrons.
\par
In 1969, Feynman observed that a fast-moving proton can be regarded
as a collection of many ``partons'' whose properties appear to be
quite different from those of the quarks~[14].  For example,
the number of quarks inside a static proton is three, while the number
of partons in a rapidly moving proton appears to be infinite.  The
question then is how the proton looking like a bound state of quarks
to one observer can appear different to an observer in a different
Lorentz frame?  Feynman made the following systematic observations.

\begin{itemize}

\item[a.]  The picture is valid only for protons moving with
  velocity close to that of light.

\item[b.]  The interaction time between the quarks becomes dilated,
   and partons behave as free independent particles.

\item[c.]  The momentum distribution of partons becomes widespread as
   the proton moves fast.

\item[d.]  The number of partons appear to be infinite or much larger
    than that of quarks.

\end{itemize}

\noindent Because the proton is believed to be a bound state of two
or three quarks, each of the above phenomena appears as a paradox,
particularly b) and c) together.

\begin{figure}[thb]
\centerline{\includegraphics[scale=0.4]{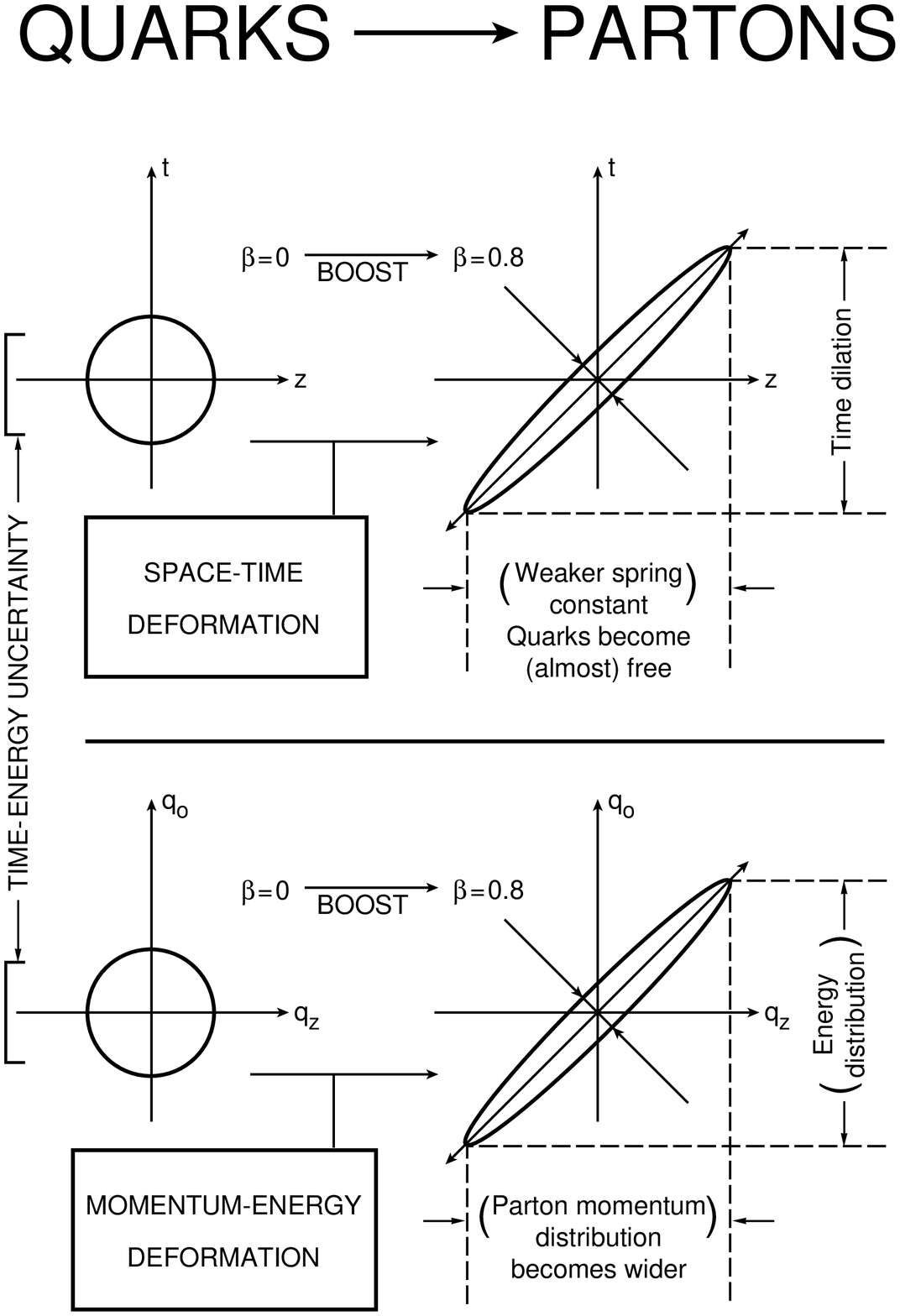}}
\caption{Lorentz-squeezed space-time and momentum-energy wave
functions.  As the proton's speed approaches that of light, both
wave functions become concentrated along their respective positive
light-cone axes.  These light-cone concentrations lead to Feynman's
parton picture.}\label{parton}
\end{figure}
\par

\begin{table}[thb]
\caption{Massive and massless particles in one package.  Einstein
unified the energy-momentum relation for slow (massive) and fast (massless)
particles with one Lorentz-covariant formula.  Likewise, can the quark
model and the parton model can be combined into one Lorentz-covariant?
The answer is YES.}\label{einpar}
\vspace{3mm}
 \begin{center}
\begin{tabular}{lccc}
\hline
{}&{}&{}&{}\\
{} & Massive & Lorentz & Massless \\
{} & Slow  & Covariance & Fast \\[2mm]\hline
{}&{}&{}&{}\\
Energy- & $E =$   & Einstein's & {} \\
Momentum &  $p^{2}/2m$ & $ E = [p^{2} + m^{2}]^{1/2}$ & $E = p$
\\[4mm]\hline
{}&{}&{}&{}\\
Relativistic & {} & One  &  {} \\[-1mm]
Extended & Quark Model & Covariant  & Parton Model\\ [-1mm]
Particles & {} & Theory &{} {} \\[4mm]\hline
\end{tabular}
\end{center}
\end{table}

\par
In order to resolve this paradox, let us consider the momentum-energy
wave function for this two-quark system.
If we let the quarks have the four-momenta $p_{a}$ and $p_{b}$, it is
possible to construct two independent four-momentum
variables~[8]
\begin{equation}
P = p_{a} + p_{b} , \qquad q = \sqrt{2}(p_{a} - p_{b}) ,
\end{equation}
where $P$ is the total four-momentum.  It is the proton four-momentum.
\par
The variable $q$ measures the four-momentum separation between
the quarks.  Their light-cone variables are
\begin{equation}\label{conju}
q_{u} = (q_{0} + q_{z})/\sqrt{2} ,  \qquad
q_{v} = (q_{0} - q_{z})/\sqrt{2} .
\end{equation}
The resulting momentum-energy wave function is
\begin{equation}\label{phi}
\phi_{\eta }(q_{z},q_{0}) =
\exp\left[-{1\over 2}\left(e^{-2\eta}q_{u}^{2} +
e^{2\eta}q_{v}^{2}\right)\right] .
\end{equation}
Because we are using here the harmonic oscillator, the mathematical
form of the above momentum-energy wave function is identical to that
of the space-time wave function.  The Lorentz squeeze properties of
these wave functions are also the same.  This aspect of the squeeze
has been exhaustively discussed in the
literature~[6,26,27].
\par
When the proton is at rest with $\eta = 0$, both wave functions
behave like those for the static bound state of quarks.  As $\eta$
increases, the wave functions become continuously squeezed until
they become concentrated along their respective positive
light-cone axes.  Let us look at the $z$-axis projection of the
space-time wave function.  Indeed, the width of the quark distribution
increases as the proton's speed approaches that of the speed of
light.  The position of each quark appears widespread to the observer
in the laboratory frame, and the quarks appear like free particles.
\par
The momentum-energy wave function is just like the space-time wave
function, as is shown in Fig.~\ref{parton}.  The longitudinal momentum
distribution becomes wide-spread as the proton's speed approaches the
velocity of light.  This is in contradiction with our expectation from
non-relativistic quantum mechanics that the width of the momentum
distribution is inversely proportional to that of the position wave
function.  Our expectation is that if the quarks are free, they must
have their sharply defined momenta, not a wide-spread distribution.
 \par
However, according to our Lorentz-squeezed space-time and
momentum-energy wave functions, the space-time width and the
momentum-energy width increase in the same direction as the proton
is boosted.  This is of course an effect of Lorentz covariance.
This indeed is the key to the resolution of the quark-parton
paradox~[6,26,27].

\par
Feynman's parton picture is one of the most controversial physical
models proposed in the 20th century.  The original model is valid
only in Lorentz frames where the initial proton moves with infinite
momentum.
It is gratifying to note that this model can be produced as a limiting
case of one covariant model which produces the quark model in the
frame where the proton is at rest.  We need Feynman's parton model
to complete the third row of Table~\ref{einpar}.

\vspace{-2mm}

\section{Concluding Remarks}

\vspace{-2.0mm}
Since 1973~[28], mostly with Marilyn Noz, I have been publishing
papers on constructing a model of quantum bound states in Einstein's
Lorentz-covariant world. In 1986~[6], we published a book on this
subject. Of course, we were not the first ones to study this problem.
\par
It was noted first that Dirac and Feynman made pivotal contributions.
However, they looked at the same problem differently in their papers.
It was seen in the present report that their results can become much
stronger if they are combined into one paper. During this process,
Wigner's 1939 paper~[5] plays the essential role.
\par
Here, the key word is ``harmony.'' The works of those great physicists
can be put together in harmony.  I am very happy to mention this
point in China, where the concept of harmony was formulated many
centuries ago through the philosophy of  "Taoism."
\par

As for Einstein, let us go to Table~\ref{einpar}.  This is a table on
harmony.  Observers in different Lorentz frames see things differently,
but they are in harmony.  Then, did Einstein study the oriental philosophy
of Taoism?  I do not know.
\par
However, it is well known that he studied the philosophy of Immanuel
Kant in his early years.  It is also known that his formulation of
relativity was influenced by Kant's view of the world.  Different
observers can see differently one thing which is called ``Ding an Sich''
by Kant. Thus, according to Kant, Einstein's special relativity requires
an absolute frame (Ding an Sich).  This is not what Einstein wanted.  In
Table~\ref{einpar}, there are no places for Kant's Ding an Sich.
\par

If not Kantianism, where is Einstein's philosophical base?  How can the
observers in two different Lorentz frames reconcile their differences?
The answer to this question seems to lie within the framework of
Taoism.  We have to study more along this direction~[29].

\section*{REFERENCES}

\vspace{-2mm}

\begin{itemize}

\item[[1]]
P. A. M. Dirac,  Proc. Roy. Soc. (London) {\bf A114}, 234 (1927).

\vspace{-1.9mm}
\item[[2]]
P. A. M. Dirac, Proc. Roy. Soc. (London) {\bf A183}, 284 (1945).

\vspace{-1.9mm}
\item[[3]]
P. A. M. Dirac,  Rev. Mod. Phys. {\bf 21}, 392 (1949).

\vspace{-1.9mm}
\item[[4]]
P. A. M. Dirac,  J. Math. Phys. {\bf 4}, 901 (1963).

\vspace{-1.9mm}
\item[[5]]
E. Wigner,  Ann. Math. {\bf 40}, 149 (1939).

\vspace{-1.9mm}
\item[[6]]
Y. S. Kim and M. E. Noz,  {\it Theory and Applications of the
Poincar\'e Group} (Reidel, Dordrecht, 1986).

\vspace{-1.9mm}
\item[[7]]
R. P. Feynman, Invited talk presented at the {\it 1970 Washington
meeting of the American Physical Society} (Washington, DC, U.S.A.).

\vspace{-1.9mm}
\item[[8]]
R. P. Feynman, M. Kislinger, and F. Ravndal, Phys. Rev. D {\bf 3},
2706 (1971).

\vspace{-1.9mm}
\item[[9]]
Y. S. Kim and M. E. Noz, Symmetry {\bf 3}, 16 (2011).

\vspace{-1.9mm}
\item[[10]]
 H. P. Yuen, Phys. Rev. A {\bf 13}, 2226 (1976).

\vspace{-1.9mm}
\item[[11]]
 J. R. Klauder, S. L. McCall, and B. Yurke, Phys. Rev. {\bf A 33},
 3204 (1986).

\vspace{-1.9mm}
\item[[12]]
Y. S. Kim, M. E. Noz, S. H. Oh, J. Math. Phys. {\bf 20}, 1341 (1979).

\vspace{-1.9mm}
\item[[13]]
R. P. Feynman, Phys. Rev. Lett. {\bf 23}, 1415 (1969).

\vspace{-1.9mm}
\item[[14]]
R. P. Feynman,  {\it The Behavior of Hadron Collisions at Extreme
Energies}, in {\it High Energy Collisions}, Proceedings of the
Third International Conference, Stony Brook, New York, edited by
C. N. Yang {\it et al.}, Pages 237-249 (Gordon and Breach,
New York, 1969).

\vspace{-1.9mm}
\item[[15]]
M. Gell-Mann, Phys. Lett. {\bf 13}, 598 (1964).

\vspace{-1.9mm}
\item[[16]]
R. P. Feynman, {\it Statistical Mechanics}
(Benjamin/Cummings, Reading, MA, (1972).

\vspace{-1.9mm}
\item[[17]]
 D. Han, Y. S. Kim, and M. E. Noz, Am. J. Phys. {\bf 67} 61 (1999).

\vspace{-1.9mm}
\item[[18]]
Y. S. Kim and E. P. Wigner, Phys. Lett. A {\bf 147}, 343 (1990).

\vspace{-1.9mm}
\item[[19]]
H. Yukawa, Phys. Rev. {\bf 91}, 415 (1953).

\vspace{-1.9mm}\item[[20]]
M. Markov, Suppl. Nuovo Cimento {\bf 3}, 760 (1956).

\vspace{-1.9mm}\item[[21]]
V. L. Ginzburg, V. I. Man'ko, Nucl. Phys. {\bf 74}, 577 (1965).

\vspace{-1.9mm}\item[[22]]
K. Fujimura, T. Kobayashi, M. Namiki,  Prog. Theor. Phys.
{\bf 1970}, 73 (1970)

\vspace{-1.9mm}\item[[23]]
A. L. Licht, A. Pagnamenta, Phys. Rev. D
{\bf 2}, 1150 (1970).

\vspace{-1.9mm}\item[[24]]
D. Han, Y. S. Kim, M. E. Noz, and D. Son, Phys. Rev. D {\bf 27},
   3032 (1983).

\vspace{-1.9mm}\item[[25]]
Y. S. kim and M. E. Noz, Am. J. Phys., {\bf 53}, 142 (1985).

\vspace{-1.9mm}\item[[26]]
Y. S. Kim and M. E. Noz, {\it Phys. Rev. D} {\bf 15}, 335 (1977).

\vspace{-1.9mm}\item[[27]]
Y. S. Kim, Phys. Rev. Lett. {\bf 63}, 348 (1989).

\vspace{-1.9mm}\item[[28]]
Y. S. Kim and M. E. Noz, {\it Phys. Rev. D} {\bf 8}, 3521 (1973).

\vspace{-1.9mm}\item[[29]]
Y. S. Kim, {\it Einstein, Kant, and Taoism} ArXiv
   http://arxiv.org/abs/physics/0604027 (2006).

\end{itemize}

\end{document}